\documentclass[12pt]{article}
\usepackage[totalheight = 23cm, totalwidth = 17cm]{geometry}
\newcommand{\nin}{\noindent}
\newcommand{\be}{\begin{equation}}
\newcommand{\ee}{\end{equation}}
\newcommand{\bea}{\begin{eqnarray}}
\newcommand{\eea}{\end{eqnarray}}

\newcommand{\hf}{\frac{1}{2}}
\newcommand{\nn}{\nonumber\\}

\usepackage{graphicx}

\begin{document}

\pagestyle{empty}
\begin{flushright}
{CERN-PH-TH/2006-201}\\
hep-th/0610072\\
\end{flushright}
\vspace*{5mm}
\begin{center}
{\large {\bf Non-Perturbative Formulation of Time-Dependent String Solutions}} \\
\vspace*{1cm}
{\bf Jean Alexandre$^1$}, {\bf John~Ellis$^{1,2}$} and {\bf Nikolaos E. Mavromatos$^1$} \\
\vspace{0.3cm}
$^1$ Department of Physics, King's College London, London  WC2R 2LS, England \\
$^2$ Theory Division, Physics Department, CERN, CH-1211 Geneva 23, Switzerland \\

\vspace*{2cm}
{\bf ABSTRACT} \\ 
\end{center}
\vspace*{5mm}
\noindent
We formulate here a new 
world-sheet renormalization-group technique 
for the bosonic string, which is non-perturbative 
in the Regge slope $\alpha^{'}$ and based on a functional method for controlling the quantum 
fluctuations, whose magnitudes are scaled by the value of $\alpha^{'}$. 
Using this technique we exhibit, in addition to the
well-known linear-dilaton cosmology, a
new, non-perturbative time-dependent background solution. Using the
reparametrization invariance of the string S-matrix, we demonstrate that this
solution is conformally invariant to ${\cal O}(\alpha^{'})$, and we give a heuristic
inductive argument that conformal invariance can be maintained to all orders in $\alpha^{'}$.
This new time-dependent string solution may be applicable to primordial cosmology or to
the exit from linear-dilaton cosmology at large times.
\vspace*{5cm}
\noindent

\begin{flushleft} CERN-PH-TH/2006-201 \\
\end{flushleft}
\vfill\eject

\setcounter{page}{1}
\pagestyle{plain}

\pagebreak

\section{Introduction}

The understanding of possible time-dependent string backgrounds is crucial for the
development of string cosmology. A first step in this direction was the solution with
a dilaton depending linearly on time that was explored in~\cite{ABEN}. In this paper
we explore further such time-dependent configurations by developing a non-perturbative
renormalization-group technique developed originally for scalar field theories~\cite{intscalar}
and applied subsequently to other models~\cite{intQED,intplanar,Farakos}. We
recover the linear-dilaton string solution in this approach, and also discover
another possible solution which has a singularity in time. This new construction
is essentially non-perturbative, so our derivation has heuristic elements. However,
we are able to support it using inductive arguments.

Our approach originates from the alternative to Wilsonian 
renormalization flows that was proposed for a scalar model in~\cite{intscalar},
where the amplitudes of quantum fluctuations are controlled by the bare mass of the field.
As this mass goes to infinity, quantum fluctuations are frozen and the system becomes classical.
On the other hand, as the bare mass decreases the quantum fluctuations grow,
and the system becomes dressed. The resulting evolution equation for the quantum theory is exact,
and provides a resummation to all orders in
$\hbar$. At the one-loop level, it recovers the well-known perturbative results but, 
beyond one loop, the gradient expansion used in this approach differs from the loop expansion.
The same approach has subsequently been applied successfully to QED~\cite{intQED},
to different (2+1)-dimensional models~\cite{intplanar} and to a Yukawa theory in five dimensions~\cite{Farakos}.

We propose here a similar functional method, in which the amplitudes of quantum fluctuations are
controlled by the amplitude of $\alpha^{'}$ and hence the string mass scale, instead of a bare mass which is not present in the world-sheet action. This enables us to avoid introducing a non-physical running cut-off on the world sheet, 
and yields an evolution equation for the dilaton which is exact and provides a 
non-perturbative resummation to all orders in $\alpha^{'}$. We still need a world-sheet cut-off
for our evolution equation, but this is fixed and can be absorbed by a rescaling
of the target space  coordinates. The new time-dependent string
solution that we find using this approach satisfies conformal-invariance conditions that we
solve explicitly at the first order in $\alpha^{'}$, but we give an inductive argument that they
are satisfied to all orders in $\alpha^{'}$. The new solution has a power-law singularity in
the metric scale factor and a logarithmic singularity in the dilaton.

We start in Section 2 by deriving the exact evolution equation for the quantum dilaton
with the amplitude of $\alpha^{'}$,
and look for exactly marginal solutions, i.e., $\alpha^{'}$-independent configurations.
We check that one such fixed-point configuration is the well-known linear dilaton/flat metric configuration~\cite{ABEN},
which was to be expected, since this configuration does not generate quantum fluctuations. 
We also discover a family of non-trivial fixed-point configurations, for which 
the target space metric is conformally flat and
the metric scale factor is proportional to the second derivative of the dilaton with respect to $X^0$.

We examine the Weyl invariance conditions in Section 3 where we show that,
although the beta functions cannot vanish order by order, there exists
one specific configuration in the family of fixed points for which the different beta functions display homogeneous dependences on 
$X^0$. This feature suggests  that this configuration has a non-perturbative nature. If the metric of this 
specific fixed point is then chosen to be proportional to $\alpha^{'}$, 
the usual expansion of the beta functions
in powers of $\alpha^{'}$ is not valid, and all the terms in the $\alpha^{'}$ expansion have to be taken into account.
We then demonstrate, using a redefinition of the dilaton and graviton fields, that it is always possible to
cancel the beta functions corresponding to Weyl invariance, so that this solution is
conformally invariant. We exhibit this property explicitly at one-loop order, and give a
heuristic inductive argument for its validity to all orders.  

Section 4 makes the connection with Wilsonian flows, using an infinitesimal renormalization-group
technique. We show that the fixed-point configuration is an infrared-stable fixed point with regard to 
Wilsonian renormalization flows. 

We examine the
cosmological properties of this string configuration in Section 5, exhibiting the singularities in the
metric scale factor and the dilaton field. Some technical aspects of our approach are outlined in two Appendices.

\section{Controlling the Amplitudes of Quantum Fluctuations}

We consider a spherical world sheet with a curvature scalar $R^{(2)}$. 
The bare action of the $\sigma$ model for the bosonic string in graviton and dilaton  
backgrounds reads
\be
S=\frac{1}{4\pi}\int d^2\xi\sqrt{\gamma}\left\{\lambda\gamma^{ab}\eta_{\mu\nu}\partial_a X^\mu\partial_b X^\nu
+R^{(2)}\phi_B(X^0)\right\},
\ee
where $\eta_{\mu\nu}$ is the flat Minkowski target-space metric. Motivated by cosmology,
we assume that the  bare dilaton field $\phi_B$ is a function
of the time coordinate only. 
The parameter $\lambda$ interpolates for $1/\alpha^{'}$ and can be used to
control the magnitude of the quantum fluctuations, 
which are proportional to $\alpha^{'}$. It parametrizes the quantum theory
described by the effective action, i.e., the proper-graph-generating functional $\Gamma_\lambda$,
which is defined in Appendix A. The range of values for $\lambda$ is $[1/\alpha^{'};\infty[$: 
\begin{itemize}
\item $\lambda\to\infty$ corresponds to $\alpha^{'}\to 0$ and therefore to a classical theory:
the kinetic term dominates over the bare dilaton $\phi_B$ term, and the theory is free;
\item $\lambda\to 1/\alpha^{'}$ generates to the full quantum theory: the interactions arising from $\phi_B$ are 
gradually switched on as $\lambda$ decreases from $\infty$ to $1/\alpha^{'}$.
\end{itemize}
We seek in this paper a $\lambda$-independent configuration of the bosonic string, which 
therefore, by definition, is non-perturbative since it is independent of the strength of $\alpha^{'}$.

\vspace{.5cm}

We derive in Appendix A the exact evolution equation for the proper-graph-generating functional
$\Gamma_\lambda$ with $\lambda$, which reads 
\bea\label{evolG}
\dot\Gamma_\lambda&=&
\frac{1}{4\pi}\int d^2\xi\sqrt{\gamma}\gamma^{ab}\eta_{\mu\nu}\partial_a X^\mu\partial_b X^\nu\\
&&+\frac{\eta_{\mu\nu}}{4\pi}\mbox{Tr}\left\{\gamma^{ab}
\frac{\partial}{\partial\xi^a}\frac{\partial}{\partial\zeta^b}
\left(\frac{\delta^2\Gamma_\lambda}{\delta X^\mu(\zeta)\delta X^\nu(\xi)}\right)^{-1}\right\},
\nonumber
\eea
where a dot over a letter denotes a derivative with respect to $\lambda$.
In eq.(\ref{evolG}), the symbol of the trace (defined in eq.(\ref{trace}) of Appendix A) 
contains the quantum corrections to $S$. In order to obtain physical information on the system,
eq.(\ref{evolG}) should in principle be integrated from $\lambda=\infty$ to $\lambda=1/\alpha^{'}$, 
which is the appropriate regime of the full quantum theory. 
An important remark is in order, so as to emphasize the difference from
Wilsonian flow and to avoid confusion:
the bare kinetic term {\it does not} play the role of a regulator. We need a
regulator (to be provided by a 
fixed cut-off $\Lambda$ in this case) so as to define our flow with $\lambda$.
The latter is therefore well defined, for a {\it fixed} regulator $\Lambda$.

We now derive the evolution equation for the 
evolution of the quantum dilaton with $\lambda$.
For this, one must have knowledge of the functional dependence
of $\Gamma_\lambda$ on the quantum fields. This can be achieved by 
means of a gradient-expansion approximation, which assumes that,
for any value of $\lambda$, $\Gamma$ takes the form:
\bea\label{gradexp}
\Gamma_\lambda&=&\frac{1}{4\pi}
\int d^2\xi\sqrt{\gamma}\left\{\gamma^{ab}\kappa_\lambda(X^0)\partial_a X^0\partial_b X^0\right.\nn
&&~~~~~~~~~~~\left.+\gamma^{ab}\tau_\lambda(X^0)\partial_a X^j\partial_b X^j+R^{(2)}\phi_\lambda(X^0)\right\},
\eea
where $\kappa_\lambda,\tau_\lambda$ are $\lambda$-dependent functions of $X^0$ which
are different for the time ($X^0$) and space ($X^j$) coordinates, since the respective 
quantum fluctuations are different.

We consider the limit where the radius of the world sheet goes to infinity, but keeping the curvature
scalar $R^{(2)}$ finite.
We show then in Appendix A that the Ansatz (\ref{gradexp}), when plugged into the exact
evolution equation (\ref{evolG}), leads to:
\be\label{equa1}
\dot\phi=-\frac{\Lambda^2}{2R^{(2)}}\left(\frac{1}{\kappa}+\frac{D-1}{\tau}\right)+
\frac{\phi^{''}}{4\kappa^2}\ln\left(1+\frac{2\Lambda^2\kappa}{R^{(2)}\phi^{''}}\right)\nonumber,
\ee
where $\Lambda$ is the world-sheet ultraviolet cut-off, 
and a prime denotes a derivative with respect to $X^0$. 
After the redefinition $X^j\to\sqrt{D-1}X^j$ of the space coordinates of the string,
one can see that the evolution equation (\ref{equa1}) is satisfied by the well-known 
flat metric/linear dilaton configuration~\cite{ABEN}:
\be\label{linear}
\kappa=1=-\tau,~~~~~~~~\phi(X^0)=QX^0, 
\ee
which shows that the latter solution is exactly marginal with respect to the flows in $\lambda$.

\vspace{0.5cm}

We now show that the evolution equation (\ref{equa1}) has, besides the known configuration (\ref{linear}), 
another solution that is also exactly marginal with respect to the $\lambda$ flow. 
We consider a configuration with $\kappa(X^0)=F\phi^{''}(X^0)$, where $F$ is a constant. For such a 
configuration, the evolution equation (\ref{equa1}) reads
\be
\kappa\dot\phi=-\frac{\Lambda^2}{2R^{(2)}}\left(1+(D-1)\frac{\kappa}{\tau}\right)
+\frac{1}{4F}\ln\left(1+\frac{2\Lambda^2F}{R^{(2)}}\right),
\ee
and one can see that it is possible to have a $\lambda$-independent solution: $\dot\phi=0$, if 
\be
\label{condition}
\frac{\kappa}{\tau}=-\frac{1}{D-1}+\frac{R^{(2)}}{2(D-1)\Lambda^2F}
\ln\left(1+\frac{2\Lambda^2F}{R^{(2)}}\right)
=-c^2,
\ee
where we have taken the negative sign corresponding to Minkowski signature, as is appropriate for large cut-off $\Lambda$, 
in which case the ratio $\kappa/\tau$ is necessarily negative, and $c$ is a positive constant. 
After the redefinition $X^j\to cX^j$ of the space coordinates of the string,
the condition (\ref{condition}) shows that the target-space metric is conformally flat, and the
non-trivial $\lambda$-independent solution of eq.(\ref{equa1}) is such that
\be\label{nontrivial}
g_{\mu\nu}(X^0)\propto\phi^{''}(X^0)\eta_{\mu\nu}.
\ee
In the next Section we display a more precise functional dependence for $\phi$, using 
the Weyl invariance conditions.

\section{Conformal Properties of the New Time-Dependent Solution}

To first order in $\alpha^{'}$, the beta functions for the bosonic world-sheet
$\sigma$-model theory in graviton and dilaton backgrounds are \cite{metsaev}:
\bea\label{weylconditions}
\beta_{\mu\nu}^g&=&R_{\mu\nu}+2\nabla_\mu\nabla_\nu\phi
+\frac{\alpha^{'}}{2}R_{\mu\lambda\rho\sigma}R_\nu^{~~\lambda\rho\sigma}+{\cal O}(\alpha^{'})^2,\nn
\beta^\phi&=&\frac{D-26}{6\alpha^{'}}-\frac{1}{2}\nabla^2\phi+\partial^\rho\phi\partial_\rho\phi
+\frac{\alpha^{'}}{16}R_{\mu\rho\nu\sigma}R^{\mu\rho\nu\sigma}+{\cal O}(\alpha^{'})^2.
\eea
We consider first the tree-level beta functions for a configuration satisfying the 
condition $\kappa(X^0)\propto\phi^{''}(X^0)$, with the power-law dependence
\bea\label{ansatz}
\phi^{'}(X^0)&=&\phi_0(X^0)^n,\nn
\kappa(X^0)&=&\kappa_0(X^0)^{n-1}.
\eea
In this case, we obtain (see Appendix B for details):
\bea
\beta^g_{00}&=&\frac{D-1}{2}\frac{n-1}{(X^0)^2}+(n+1)\phi_0(X^0)^{n-1}+{\cal O}(\alpha^{'}),\nn
\beta^g_{jk}&=&\delta_{jk}\left(\frac{(n-1)(D-2)-2}{4(X^0)^2}-\phi_0(X^0)^{n-1}\right)+{\cal O}(\alpha^{'}),\nn
\beta^\phi&=&\frac{D-26}{6\alpha^{'}}-\frac{\phi_0}{4\kappa_0}\Big(n+1+(n-1)(D-1)\Big)
+\frac{\phi_0^2}{\kappa_0}(X^0)^{n+1}+{\cal O}(\alpha^{'}).
\eea
We observe that, for $n=-1$, each beta function is homogeneous, and we have  
\bea
\beta^g_{00}&=&-\frac{D-1}{(X^0)^2}+{\cal O}(\alpha^{'}),\nn
\beta^g_{jk}&=&\delta_{jk}\frac{D-1+2\phi_0}{(X^0)^2}+{\cal O}(\alpha^{'}),\nn
\beta^\phi&=&\frac{D-26}{6\alpha^{'}}+\phi_0\frac{D-1+2\phi_0}{2\kappa_0}+{\cal O}(\alpha^{'}).
\eea
We see that it is not possible for $\beta^g_{00}$ to vanish at the tree level. 

An important remark is in order, though: 
for the specific choice $n=-1$, the next order of each beta function is homogeneous with the tree-level
term (see Appendix B for details):
\bea
R_{0\mu\nu\rho}R_0^{~~\mu\nu\rho}&=&\frac{3(D-1)}{\kappa_0(X^0)^2},\nn
R_{j\mu\nu\rho}R_k^{~~\mu\nu\rho}&=&-\delta_{jk}\frac{2D-1}{\kappa_0(X^0)^2},\nn
R_{\mu\nu\rho\sigma}R^{\mu\nu\rho\sigma}&=&\frac{2(D^2-1)}{\kappa_0^2}.
\eea
This is actually valid to all orders, 
as can readily be seen using 
the fact that at each order in $\alpha^{'}$ the 
Weyl anomaly coefficients of the dilaton and graviton backgrounds consist of 
terms that involve products of appropriate powers of Riemann tensors, 
dilaton (covariant) derivatives, and the necessary contravariant 
metric tensors $g^{\mu\nu} \propto \phi''(X^0) \eta^{\mu\nu}$ (c.f., (\ref{nontrivial})) 
for the appropriate contractions. Inspection of such terms then
reveals, in a similar way to the ${\cal O}(\alpha^{'})$ terms above, 
homogeneous $1/(X^0)^2$ behaviours for cosmological backgrounds 
satisfying (\ref{ansatz}). 

We then have, for the Ansatz (\ref{ansatz}) with $n=-1$,
\bea
\beta^g_{00}&=&\frac{1}{(X^0)^2}\sum_{m=0}^\infty \xi_m\left(\frac{\alpha^{'}}{\kappa_0}\right)^m,\nn
\beta^g_{jk}&=&\frac{\delta_{jk}}{(X^0)^2}\sum_{m=0}^\infty \zeta_m\left(\frac{\alpha^{'}}{\kappa_0}\right)^m,\nn
\beta^\phi&=&\frac{1}{\alpha^{'}}\sum_{m=0}^\infty \eta_m\left(\frac{\alpha^{'}}{\kappa_0}\right)^m,
\eea
where $\xi_m,\zeta_m,\eta_m$ are $\alpha^{'}$-independent coefficients. As a consequence,
for $\kappa_0$ of the same order as $\alpha^{'}$, the expansion of the beta functions in $\alpha^{'}$ 
is no longer valid. 

The next step is to argue that the configuration 
\bea\label{config}
\phi(X^0)&=&\phi_0\ln(X^0),\nn
\kappa(X^0)&=&\frac{\alpha^{'}A}{(X^0)^2},
\eea 
where we write $\kappa_0=\alpha^{'}A$, 
may satisfy conformal invariance at a non-perturbative level.
We exploit the fact, well-known in string theory, that at higher orders in $\alpha^{'}$ the beta functions  are not fixed uniquely, but can be changed by making local
field redefinitions~\cite{metsaev}: $g_{\mu\nu}\to\tilde g_{\mu\nu}$ and $\phi\to\tilde\phi$,
which leave the (perturbative) string S-matrix amplitudes invariant.
This possibility of field redefinition enables us to maintain conformal invariance to
all orders in $\alpha^{'}$.

We illustrate this possibility with an explicit calculation to first order in $\alpha^{'}$. 
In our case, since we keep the target-space metric
conformally flat, the redefinition of the metric must be such that $\tilde g_{\mu\nu}$
is proportional to $g_{\mu\nu}$, and thus we can consider the following redefinitions:
\bea\label{redef}
\tilde g_{\mu\nu}&=&g_{\mu\nu}+\alpha^{'}g_{\mu\nu}
\left(b_1R+b_2\partial^\rho\phi\partial_\rho\phi+b_3\nabla^2\phi\right),\nn
\tilde\phi&=&\phi+\alpha^{'}
\left(c_1R+c_2\partial^\rho\phi\partial_\rho\phi+c_3\nabla^2\phi\right),
\eea
where $b_1,b_2,b_3,c_1,c_2,c_3$ are constants and $R,\nabla$ 
corresponds to the metric $g_{\mu\nu}$.
In the case of the configuration (\ref{config}), we have (see Appendix B for details):
\bea\label{gtilde}
\tilde g_{\mu\nu}&=&g_{\mu\nu}+\frac{g_{\mu\nu}}{A}\Big(-b_1(D-1)^2+b_2\phi_0^2-b_3(D-1)\phi_0\Big)=(1+B)g_{\mu\nu},\nn
\tilde\phi&=&\phi+\frac{1}{A}\Big(-c_1(D-1)^2+c_2\phi_0^2-c_3(D-1)\phi_0\Big)=\phi+C,
\eea
where $B,C$ are constants linear in $b_1,b_2,b_3,c_1,c_2,c_3$.
Therefore, the redefinitions (\ref{redef}) consist of adding a constant to the dilaton and rescaling
the metric, thus not changing the functional dependence of the configuration (\ref{config}).

We now examine the changes in the beta functions after the field redefinitions 
(\ref{redef}). The new beta functions 
$\beta_{\mu\nu}^g\to\tilde\beta_{\mu\nu}^g$ and $\beta^\phi
\to\tilde\beta^\phi$ are obtained via the appropriate Lie derivatives in theory
space~\cite{metsaev}, as appropriate to the vector nature of the
$\beta^i$ functions in this space:
\bea\label{betatildeg}
\tilde\beta_{\mu\nu}^g-\beta_{\mu\nu}^g
&=&\int(\tilde g_{\rho\sigma}-g_{\rho\sigma})\frac{\delta\beta_{\mu\nu}^g}{\delta g_{\rho\sigma}}
+\int(\tilde\phi-\phi)\frac{\delta\beta_{\mu\nu}^g}{\delta\phi}\nn
&&-\int\beta_{\rho\sigma}^g\frac{\delta(\tilde g_{\mu\nu}-g_{\mu\nu})}{\delta g_{\rho\sigma}}
-\int\beta^\phi\frac{\delta(\tilde g_{\mu\nu}-g_{\mu\nu})}{\delta\phi},
\eea
and
\bea\label{betatildephi}
\tilde\beta^\phi-\beta^\phi
&=&\int(\tilde g_{\rho\sigma}-g_{\rho\sigma})\frac{\delta\beta^\phi}{\delta g_{\rho\sigma}}
+\int(\tilde\phi-\phi)\frac{\delta\beta^\phi}{\delta\phi}\nn
&&-\int\beta_{\rho\sigma}^g\frac{\delta(\tilde\phi-\phi)}{\delta g_{\rho\sigma}}
-\int\beta^\phi\frac{\delta(\tilde\phi-\phi)}{\delta\phi}.
\eea
After long but straightforward computations, we obtain
\bea
\tilde\beta^g_{00}&=&\frac{1}{(X^0)^2}\Bigg\{1-D+2C+3BD\left(1-D-\phi_0+\frac{5(D-1)}{A}\right)\\
&&~~~~-\frac{D-1}{A}\left(\frac{D+2}{A}-D-2\phi_0\right)\left(b_1(D-1)(6D-13)+\frac{3}{2}b_3\phi_0(D-2)\right)\nn
&&~~~~-\left(\frac{D-26}{6}+\phi_0\frac{D-1+2\phi_0}{2A}+\frac{D^2-1}{8A^2}\right)\Big(2b_2+(2-D)b_3\Big)\Bigg\},\nn
\tilde\beta^g_{jk}&=&\frac{\delta_{jk}}{(X^0)^2}\Bigg\{D-1+2\phi_0+2C+BD\left(15-6D-3\phi_0+\frac{14D-43}{A}\right)\nn
&&~~~~-\frac{D-1}{A}\left(2\phi_0+D-\frac{D+2}{A}\right)\left(b_1(D-1)(6D-13)+\frac{3}{2}b_3\phi_0(D-2)\right)\nn
&&~~~~+\left(\frac{D-26}{6}+\phi_0\frac{D-1+2\phi_0}{2A}+\frac{D^2-1}{8A^2}\right)\Big(2b_2+(2-D)b_3\Big)\Bigg\},\nn
\tilde\beta^\phi&=&\frac{1}{\alpha^{'}}\Bigg\{\left(\frac{D-26}{6}+\phi_0\frac{D-1+2\phi_0}{2A}+\frac{D^2-1}{8A^2}\right)
\left(1-\frac{c_3D-2c_2\phi_0}{A}\right)\nn
&&~~~~+\frac{BD}{A}\left(\frac{\phi_0}{4}(3D-4)-\phi_0^2+\frac{4}{A}(D-1)(D+10)\right)
-\frac{C}{A}\left(\frac{D}{2}+2\phi_0\right)\nn
&&~~~~-\frac{D-1}{A^2}\left(\frac{D+2}{A}-D-2\phi_0\right)
\left(3c_1D(D-1)-c_2\phi_0^2+\frac{c_3}{2}(3D-4)\phi_0\right)\Bigg\}.\nonumber
\eea
We observe that, for any dimension $D$ and any dilaton amplitude $\phi_0$, one can always find a set of parameters
$b_1,b_2,b_3,c_1,c_2,c_3$ such that all the three beta functions $\tilde\beta^g_{00},\tilde\beta^g_{jk},\tilde\beta^\phi$ vanish,
since the latter are homogeneous in $X^0$.

In order to check that there are indeed sets of these 
redefinition parameters which cancel the
the beta functions at first order in $\alpha^{'}$, we consider the special case where $b_1=b_3=c_2=0$, 
so that we are left with a linear system
of three equations to solve for the three variables $b_2,c_1,c_3$. The determinant of this system is
\bea
\mbox{det}(A,D,\phi_0)&=&-\frac{(D-1)^2}{72A^6}\Big(-24D^2A\phi_0-336DA\phi_0^2+84D^2\phi_0+36D^3\phi_0\nn
&&-24D\phi_0-36D^3A\phi_0+216D^2A\phi_0^2+60DA\phi_0+96A\phi_0\nn
&&-96\phi_0-4D^2A^2+104DA^2-3D^3+3D\Big)\nn
&&\times\Big(-72DA\phi_0^2+18D^2A\phi_0^2+6D^2\phi_0^2+168D\phi_0^2\nn
&&-4DA^2+104A^2-12DA\phi_0+12A\phi_0-24A\phi_0^2-3D^2+3\Big).
\eea
This does not vanish in the general case, so the set of three simultaneous equations
is not degenerate, and we can indeed find values of the redefinition parameters $b_2,c_1,c_3$
that satisfy conformal invariance. This is valid for any dimension $D$ and dilaton amplitude $\phi_0$,
if the metric amplitude $A$ is chosen in a way such that the determinant does not vanish. The determinant is a
quartic function of $A$, so that there are at most four discrete values of $A$ which are not allowed in this specific 
field redefinition, for given values of $D,\phi_0$. 
However, these four values do not have any physical significance {\it a priori}, and any
other choice of field redefinition, e.g., $b_2=c_1=c_3=0$, would correspond to
different `forbidden' values for $A$.
As discussed in Section 5, the specific values $D=4,\phi_0=-1$ are particularly
interesting, since they lead to
a four-dimensional Minkowski target space-time. In this case, the determinant is 
\be
\mbox{det}(A,4,-1)=\frac{1}{2A^6}\left(909-1164A-58A^2\right)\left(723+12A+88A^2\right),
\ee
which vanishes for only two values of $A$. We emphasize
again that these particular values have no physical relevance.

This analysis took into account only the first non-trivial order in $\alpha^{'}$, whereas all the 
higher orders should also be taken into account. However, this first-order analysis
provides the basis for an inductive argument. If conformal invariance is satisfied at 
order $n$ in $\alpha^{'}$, as in the first-order case worked out above, there are always
enough parameters in the redefinitions of the metric and dilaton at the next order,
leaving the string configuration unchanged, which enable the beta functions to vanish and hence 
conformal invariance to be satisfied at the 
next order $n+1$ in $\alpha^{'}$.

It is instructive to compare and contrast the approach used here with the $\epsilon$ expansion 
used in condensed-matter problems. There, as here, the $\beta$ function contains a non-vanishing 
term at the tree level, given in that case by dimensional analysis and here by the 
${\cal O}((\alpha^{'})^0)$ term. There, as here, this term is cancelled by the one-loop term 
to yield a non-trivial infrared fixed point. In our case, this cancellation is made possible 
by the reparametrization invariance of the string S-matrix, and should hold to all orders in 
$\alpha^{'}$. In the $\epsilon$ expansion, dimensional regularization is used and the 
non-trivial fixed-point is modified perturbatively in higher loop orders. 
These provide corrections that are higher order in $\epsilon$, notionally a small 
parameter whose physical value is usually $\epsilon = 1$ or 2. Nevertheless, the $\epsilon$ 
expansion gives useful quantitative results. In our case, there is no small expansion parameter, 
but the qualitative conclusions drawn from this one-loop analysis should hold in higher orders.

\section{Wilsonian Interpretation of the New Time-Dependent Solution}

We now exhibit the Wilsonian properties of the configuration (\ref{config}), 
using the exact renormalization method of~\cite{WH}.
We consider an initial bare theory defined on the world sheet of the string, with cut-off $\Lambda$,
as described above. The
effective theory defined by the action $S_k$ at the scale $k$ is derived by integrating the ultraviolet 
degrees of freedom from $\Lambda$ to $k$. The idea of exact renormalization methods is to 
perform this integration infinitesimally, from $k$ to $k-\delta k$, 
which leads to an exact evolution equation 
for $S_k$ in the limit $\delta k<<k$. The procedure was detailed in~\cite{WH}, and here we reproduce 
only the main steps for clarity and completeness. Note that we consider here a sharp cut-off, 
which is possible only if we consider 
the evolution of the dilaton, as explained now.

We consider a Euclidean world-sheet metric, and we assume that, 
for each value of the energy scale $k$, the Euclidean action $S_k$ has the form  
\be\label{Euclansatz}
S_k=\frac{1}{4\pi\alpha^{'}}\int d^2\xi\sqrt\gamma
\Big\{\gamma^{ab}\kappa_k(X^0)\delta_{\mu\nu}\partial_a X^\mu\partial_b X^\nu
+\alpha^{'}R^{(2)}\phi_k(X^0)\Big\},
\ee
where $R^{(2)}$ is the curvature scalar of the spherical world sheet.
The integration of the ultraviolet degrees of freedom is implemented in the following way. We write the 
dynamical fields $X^\mu$ as $X^\mu=x^\mu+y^\mu$, where the $x^\mu$ are the infrared fields with 
non-vanishing Fourier components 
for $|p|\le k-\delta k$, and the $y^\mu$ are the degrees of freedom to be integrated out, with 
non-vanishing Fourier components for $k-\delta k<|p|\le k$ only. 
An infinitesimal step of the renormalization group transformation 
reads (we take the limit of a flat world-sheet metric, keeping $R^{(2)}$ finite):
\bea\label{transfo}
&&\exp\left(-S_{k-\delta k}[x]+S_k[x]\right)\\
&=&\exp\left(S_k[x]\right)\int {\cal D}[y]\exp\left(-S_k[x+y]\right)\nn
&=&\int{\cal D}[y]\exp\left(-\int_k \frac{\delta S_k[x]}{\delta y^\mu(p)}y^\mu(p)
-\hf\int_k\int_k\frac{\delta^2S_k[x]}{\delta y^\mu(p)\delta y^\nu(q)}y^\mu(p)y^\nu(q)\right),\nn
&&~~~~~~~~~~~~~~+\mbox{higher orders in}~\delta k, \nonumber
\eea
where $\int_k$ represents the integration over Fourier modes for $k-\delta k<|p|\le k$.
Higher-order terms in the expansion of the action are indeed of higher order in $\delta k$, since each integral 
involves a new factor of $\delta k$. The only relevant terms are of first and second order in 
$\delta k$ \cite{WH}, which are at most quadratic in the dynamical 
variable $y$, and therefore lead to a Gaussian integral. We then have
\bea\label{evolS}
\frac{S_k[x]-S_{k-\delta k}[x]}{\delta k}&=&\frac{\mbox{Tr}_k}{\delta k}\left\{\frac{\delta S_k[x]}{\delta y^\mu(p)}
\left(\frac{\delta^2S_k[x]}{\delta y^\mu(p)\delta y^\nu(q)}\right)^{-1}
\frac{\delta S_k[x]}{\delta y^\nu(q)}\right\}\nn
&&-\frac{\mbox{Tr}_k}{2\delta k}\left\{\ln\left(\frac{\delta^2S_k[x]}{\delta y^\mu(p)\delta y^\nu(q)}\right)\right\}
+{\cal O}(\delta k),
\eea
where the trace Tr$_k$ is to be taken in the shell of thickness $\delta k$, and is therefore
proportional to $\delta k$.

We are interested in the evolution equation of the dilaton, for which it is sufficient to 
consider a constant infrared configuration $x^\mu$~\footnote{This is why a sharp cut-off can be used: 
the singular terms that could arise from the $\theta$ function, characterizing the sharp cut-off,
are not present, since the derivatives of the infrared field vanish.}.
In this situation, the first term on the right-hand
side of eq.(\ref{evolS}), which 
is a tree-level term, does not contribute: $\delta S_k/\delta y^\mu(p)$ is proportional 
to $\delta^2(p)$, and thus has no overlap with the domain of integration $|p|=k$.
We are therefore left with the second term, which arises from quantum fluctuations, and
the limit $\delta k\to 0$ gives, with the Ansatz (\ref{Euclansatz}),
\be\label{evolequa}
R^{(2)}\partial_k\phi_k(x^0)=
-k\ln\left(\frac{2\kappa_k(x^0)k^2+\alpha^{'}R^{(2)}\phi^{''}_k(x^0)}
{2\kappa_k(1)k^2+\alpha^{'}R^{(2)}\phi^{''}_k(1)}
\left(\frac{\kappa_k(x^0)}{\kappa_k(1)}\right)^{D-1}\right),
\ee
where a prime denotes a derivative with respect to $X^0$ and 
we chose the dilaton to vanish for $x^0=1$.
Eq.(\ref{evolequa}) provides a resummation in $\alpha^{'}$,
since the quantities in the logarithm are the running, dressed quantities. As a result, the 
evolution equation (\ref{evolequa}) is exact within the framework of the Ansatz (\ref{Euclansatz}),
and is non-perturbative.

One can easily see that a linear dilaton/flat metric configuration~\cite{ABEN}
\be
\kappa(x^0)=1~~~~~~~~\phi(x^0)=Qx^0,
\ee
where the constant $Q$ is independent of $k$, satisfies the evolution equation (\ref{evolequa}).
This is an exactly marginal configuration, independent of the 
Wilsonian scale $k$, which is expected because 
it does not generate any quantum fluctuations. 

\vspace{0.5cm}

We now come back to the configuration (\ref{config}), and look for a similar exact solution of 
the renormalization-group equation (\ref{evolequa}), in which the graviton background is given by
the same expression as in eq.(\ref{config}), but the dilaton has the form
\be
\phi_k(x^0)=\eta_k\ln(x^0),
\ee
where $\eta_k$ is a function of $k$. It is easy to see that such a configuration 
indeed satisfies eq.(\ref{evolequa}), provided that $\eta_k$ satisfies $d\eta_k/dk=2Dk/R^{(2)}$, 
and hence
\be\label{sol}
\phi_k(x^0)=\left(\phi_0+\frac{Dk^2}{R^{(2)}}\right)\ln(x^0),
\ee
where $\phi_0$ is the constant of integration.

Therefore, we have been able to find an exact solution of the renormalization-group equation (\ref{evolequa}),
which tends to the solution (\ref{config}) in the infrared limit $k\to 0$, with a vanishing derivative 
\be
\partial_k\phi_k(x^0)\to 0^{+},
\ee 
and thus we conclude that the configuration (\ref{config}) is a Wilsonian infrared-stable fixed point.

\section{Cosmological Properties of the New Time-Dependent Solution}

We now examine the physical significance of the new non-trivial fixed-point solution (\ref{config}),
and discuss briefly its cosmological implications. This leads to a value of the constant $\phi_0$ 
that appears in the configuration (\ref{config}).

The relation between the physical metric in the Einstein frame and the string metric is given 
by~\cite{ABEN}
\bea
ds^2&=&dt^2-a^2(t)dx^kdx^k\nn
&=&\kappa(x^0)\exp\left\{-\frac{4\phi}{D-2}\right\}\left(dx^0dx^0-dx^kdx^k\right),
\eea
where $a(t)$ is the scale factor of a spatially flat Robertson-Walker-Friedmann Universe,
and the $x^\mu$ are the zero modes of $X^\mu$. From the configuration (\ref{config}), we have
\be
\frac{dt}{dx^0}=\varepsilon\sqrt{|\kappa_0|} (x^0)^{-1-\frac{2\phi_0}{D-2}},
\ee
where $\varepsilon=\pm 1$, such that
\be\label{tx^0}
t=T+\sqrt{|\kappa_0|}\frac{(D-2)}{2|\phi_0|}(x^0)^{-\frac{2\phi_0}{D-2}},
\ee
where $T$ is a constant. We find then a power law for the evolution of the scale factor:
\be\label{scaleexpans}
a(t)=a_0|t-T|^{\frac{D-2}{2\phi_0}+1},
\ee
which is in general singular as $t \to T$.

In order to have a Minkowski target space, one needs
$D-2+2\phi_0=0$. As was discussed in Section 3, 
when dealing with the conformal properties of the 
configuration (\ref{config}), the choices of $D$ and $\phi_0$ are free, and lead to the 
determination of $\kappa_0$ (in a way which has not been determined yet). 
As a consequence, for a given dimension $D$, it is always 
possible to choose $\phi_0$ so that the target space is static and flat.
It may therefore find an application to the exit phase from the
linearly expanding Universe associated with the linear dilaton of~\cite{ABEN}.

We note that, in terms of the Einstein time $t$, the dilaton can be written, up to a constant, as:
\be\label{logdil}
\phi=-\frac{D-2}{2}\ln|t-T|.
\ee
We observe that, like the scale factor (\ref{scaleexpans}), the dilaton has a
singularity as $t \to T$. It would be interesting to explore the applicability of
this configuration to primordial cosmology. The sign of the expression (\ref{logdil}) for the
dilaton when $D > 2$ ensures that the string coupling is small at large times.

\section{Conclusions}

We have proposed in this paper a new non-perturbative renormalization-group technique 
for the bosonic string, based on a functional method for controlling the quantum fluctuations, whose magnitudes are scaled by the value of $\alpha^{'}$. Using this technique, we have exhibited a
new, non-perturbative time-dependent background solution. Using the
reparametrization invariance of the string S-matrix, we have demonstrated that this
solution is conformally invariant to ${\cal O}(\alpha^{'})$, and we have given a heuristic
inductive argument that conformal invariance can be maintained to all orders in $\alpha^{'}$.

This new non-perturbative time-dependent background solution has related singularities in both
the metric scale factor and the dilaton value at a specific value of the time in the Einstein frame.
A full exploration of the possible cosmological applications of this solution lies beyond the
scope of this paper, but we do note two interesting possibilities. One is that the temporal 
singularity might be relevant for primordial cosmology, i.e., the beginning of the Big Bang. The
second possible application could be to describe the exit phase from the
linearly expanding Universe associated with the linear dilaton of~\cite{ABEN}.

These are two phenomenological tasks for future work on this new non-perturbative 
time-dependent background solution. It is also desirable to explore in more detail the
formal underpinnings of the solution. In particular, it is necessary to improve on our
heuristic inductive argument that its conformal invariance may be maintained to all orders
in $\alpha^{'}$. We also note that the non-perturbative renormalization-group technique 
proposed here may have applications to other aspects of string theory.

\section*{Acknowledgements} 

The work of J.E. and N.E.M. was supported in part by the European Union through the Marie Curie Research and Training Network UniverseNet (MRTN-CT-2006-035863). 

\section*{Appendix A: A Novel Non-Perturbative World-Sheet Functional 
Renormalization Method for the Bosonic String}

We start with the bare action for the bosonic string on a Euclidean world sheet, expressed in terms of 
microscopic fields $\tilde X^\mu$ defined in the bare theory:
\be
S_B=\frac{1}{4\pi}\int d^2\xi\sqrt{\gamma}\left\{
\gamma^{ab}\lambda\eta_{\mu\nu}\partial_a\tilde X^\mu\partial_b\tilde X^\nu
+R^{(2)}\phi_B(\tilde X^0)\right\},
\ee
to which we add the source term
\be
S_S=\int d^2\xi\sqrt{\gamma}R^{(2)}\eta_{\mu\nu}V^\mu\tilde X^\nu,
\ee
in order to define the classical fields. The corresponding partition function $Z$ and 
the generating functional of the connected graphs $W$ are
related as usual:
\be
Z=\int{\cal D}[\tilde X]e^{-S_B-S_S}=e^{-W}.
\ee
The classical fields $X^\mu$ are defined by
\be
X^\mu=\frac{1}{Z}\int{\cal D}[\tilde X]\tilde X^\mu e^{-S_B-S_S}=\frac{1}{Z}\left<\tilde X^\mu\right>,
\ee
and are obtained by differentiating $W$ with respect to the source $V_\mu$:
\be\label{VX}
\frac{1}{\sqrt{\gamma_\xi}R^{(2)}_\xi}\frac{\delta W}{\delta V_\mu(\xi)}=X^\mu(\xi).
\ee
The second derivative of $W$ is then
\be
\frac{1}{\sqrt{\gamma_\zeta\gamma_\xi}R^{(2)}_\zeta R^{(2)}_\xi}\frac{\delta^2 W}{\delta V_\mu(\zeta)\delta V_\nu(\xi)}
=X^\nu(\xi)X^\mu(\zeta)-\frac{\left<\tilde X^\nu(\xi)\tilde X^\mu(\zeta)\right>}{Z}.
\ee
Inverting the relation (\ref{VX}) between $V_\mu$ and $X^\mu$, 
we then introduce the Legendre transform of $W$, namely
the functional $\Gamma$ responsible for the generation of 
proper graphs:
\be
\Gamma=W-\int d^2\xi\sqrt{\gamma}R^{(2)}V^\mu X_\mu.
\ee
The functional derivatives of $\Gamma$ are:
\bea\label{derivG}
\frac{1}{\sqrt{\gamma_\xi}R^{(2)}_\xi}\frac{\delta\Gamma}{\delta X^\mu(\xi)}&=&-V_\mu(\xi),\nn
\frac{1}{\sqrt{\gamma_\xi\gamma_\zeta}R^{(2)}_\xi R^{(2)}_\zeta}
\frac{\delta^2\Gamma}{\delta X^\nu(\zeta)\delta X^\mu(\xi)}&=&
-\left(\frac{\delta^2W}{\delta V_\nu(\zeta)\delta V_\mu(\xi)}\right)^{-1}.
\eea

\vspace{.5cm}

\nin The evolution of $W$ with $\lambda$ is given by
\bea
\dot W
&=&\frac{1}{4\pi Z}\int d^2\xi\sqrt{\gamma}\gamma^{ab}\eta_{\mu\nu}\left<
\partial_a\tilde X^\mu\partial_b\tilde X^\nu\right>\\
&=&\frac{\eta_{\mu\nu}}{4\pi Z}\mbox{Tr}\left\{\gamma^{ab}
\frac{\partial}{\partial\xi^a}\frac{\partial}{\partial\zeta^b}
\left<\tilde X^\mu(\xi)\tilde X^\nu(\zeta)\right>\right\},\nonumber
\eea
where a dot denotes a derivative with respect to $\lambda$, and the trace is
defined by
\be\label{trace}
\mbox{Tr}\{\cdot\cdot\cdot\}=
\int d^2\xi d^2\zeta\sqrt{\gamma_\xi\gamma_\zeta}\{\cdot\cdot\cdot\}\delta^2(\xi-\zeta).
\ee
We then obtain
\bea\label{evolW}
\dot W&=&\frac{1}{4\pi}\int d^2\xi\sqrt{\gamma}\gamma^{ab}\eta_{\mu\nu}\partial_a X^\mu\partial_b X^\nu\\
&&-\frac{\eta_{\mu\nu}}{4\pi}\mbox{Tr}\left\{
\gamma^{ab}\frac{\partial}{\partial\xi^a}\frac{\partial}{\partial\zeta^b}
\left(\frac{1}{\sqrt{\gamma_\xi\gamma_\zeta}R^{(2)}_\xi R^{(2)}_\zeta}
\frac{\delta^2 W}{\delta V_\mu(\zeta)\delta V_\nu(\xi)}\right)\right\}\nonumber.
\eea
Finally, the evolution of $\Gamma$ is obtained by noting 
that the independent variables of
$\Gamma$ are $X^\mu$ and $\lambda$, and that
\bea\label{dotG}
\dot\Gamma&=&\dot W+\int d^2\xi\frac{\partial W}{\partial V_\mu}
\dot V_\mu-\int d^2\xi\sqrt\gamma R\dot V_\mu X^\mu\nn
&=&\dot W.
\eea
Using eqs.(\ref{derivG}), 
(\ref{evolW}) and (\ref{dotG}),
we obtain finally 
the following evolution equation 
for $\Gamma$:
\bea\label{evolGappendix}
\dot\Gamma&=&\frac{1}{4\pi}\int d^2\xi\sqrt{\gamma}\gamma^{ab}\eta_{\mu\nu}\partial_a X^\mu\partial_b X^\nu\\
&&+\frac{\eta_{\mu\nu}}{4\pi}\mbox{Tr}\left\{\gamma^{ab}
\frac{\partial}{\partial\xi^a}\frac{\partial}{\partial\zeta^b}
\left(\frac{\delta^2\Gamma}{\delta X^\mu(\zeta)\delta X^\nu(\xi)}\right)^{-1}\right\}.\nonumber
\eea

\noindent
We now assume the following functional dependence:
\be
\Gamma=\frac{1}{4\pi}\int d^2\xi\sqrt{\gamma}\left\{\gamma^{ab}\left(\kappa_\lambda(X^0)\partial_a X^0\partial_b X^0
+\tau_\lambda(X^0)\partial_a X^k\partial_b X^k\right)+R^{(2)}\phi_\lambda(X^0)\right\}.
\ee
With the metric $\gamma^{ab}=\delta^{ab}$ and a constant configuration $X^0(\xi)=x^0$, 
the second functional derivatives of $\Gamma$ are
\bea
\frac{\delta^2\Gamma}{\delta X^0(\zeta)\delta X^0(\xi)}&=&
-\frac{\kappa}{2\pi}\Delta\delta^2(\xi-\zeta)+\frac{R^{(2)}\phi^{''}}{4\pi}\delta^2(\xi-\zeta),\nn
\frac{\delta^2\Gamma}{\delta X^j(\zeta)\delta X^k(\xi)}&=&
-\frac{\tau}{2\pi}\Delta\delta^2(\xi-\zeta)\delta_{jk},
\eea
where a prime denotes a derivative with respect to $x_0$.
The second functional derivatives of $\Gamma$ read then, in Fourier components,
\bea
\frac{\delta^2\Gamma}{\delta X^0(p)\delta X^0(q)}&=&
\frac{1}{4\pi}\left(2\kappa p^2+R^{(2)}\phi^{''}\right)\delta^2(p+q)\nn
\frac{\delta^2\Gamma}{\delta X^j(p)\delta X^k(q)}&=&
\frac{\tau p^2}{2\pi}\delta^2(p+q)\delta_{jk}.
\eea

\noindent
The area of a sphere with curvature scalar $R^{(2)}$ is $8\pi/R^{(2)}$, so
with the constant configuration $X^0=x_0$ we have
\be\label{Gammaconfig}
\Gamma=2\phi_\lambda(x_0),
\ee
and the trace appearing in eq.(\ref{evolGappendix}) is
\bea\label{TrABA}
\frac{1}{4\pi}\mbox{Tr}\{\partial\partial (\delta^2\Gamma)^{-1}\}
&=&-\int\frac{d^2p}{(2\pi)^2}\left(\frac{p^2}{2\kappa p^2+R^{(2)}\phi^{''}}
+\frac{D-1}{2\tau}\right)\frac{8\pi}{R^{(2)}}\nn
&=&-\frac{\Lambda^2}{R^{(2)}}\left(\frac{1}{\kappa}+\frac{D-1}{\tau}\right)
+\frac{\phi^{''}}{2\kappa^2}\ln\left(1+\frac{2\Lambda^2\kappa}{R^{(2)}\phi^{''}}\right).
\eea
To compute this trace, we used the fact that 
\be
\delta^2(p=0)=~\mbox{world-sheet area}~=\frac{8\pi}{R^{(2)}}.
\ee
The evolution equation for $\phi$ 
is finally obtained by putting together results
(\ref{evolGappendix}), (\ref{Gammaconfig}) and (\ref{TrABA}):
\be
\dot\phi=-\frac{\Lambda^2}{2R^{(2)}}\left(\frac{1}{\kappa}+\frac{D-1}{\tau}\right)+
\frac{\phi^{''}}{4\kappa^2}\ln\left(1+\frac{2\Lambda^2\kappa}{R^{(2)}\phi^{''}}\right).
\ee

\section*{Appendix B: Geometrical Properties of the Graviton 
and Dilaton Backgrounds}

For the target-space metric $g_{\mu\nu}(X^0)=\kappa(X^0)\eta_{\mu\nu}$, 
the non-vanishing Christofel symbols are
\be
\Gamma^j_{~0k}=\delta^j_k\frac{\kappa^{'}}{2\kappa},~~~~~~~
\Gamma^0_{~00}=\frac{\kappa^{'}}{2\kappa},~~~~~~~~
\Gamma^0_{~jk}=\delta_{jk}\frac{\kappa^{'}}{2\kappa},
\ee
so that the non vanishing covariant derivatives of the dilaton are
\bea
\nabla_0\nabla_0\phi&=&\phi^{''}-\frac{\kappa^{'}}{2\kappa}\phi^{'},\nn
\nabla_j\nabla_k\phi&=&-\delta_{jk}\frac{\kappa^{'}}{2\kappa}\phi^{'},
\eea
and the non-vanishing components of the Riemann and Ricci tensors are
\bea
R_{0j0}^{~~~~k}&=&-\delta_j^k\left(\frac{\kappa^{'}}{2\kappa}\right)^{'},\nn
R_{lkm}^{~~~~j}&=&\left(\frac{\kappa^{'}}{2\kappa}\right)^2\left(\delta_{lm}\delta^j_k-\delta_{km}\delta^j_l\right),\nn
R_{00}&=&-(D-1)\left(\frac{\kappa^{'}}{2\kappa}\right)^{'},\nn
R_{jk}&=&\delta_{jk}\left(\frac{\kappa^{'}}{2\kappa}\right)^{'}
+\delta_{jk}(D-2)\left(\frac{\kappa^{'}}{2\kappa}\right)^2.
\eea
We then have
\bea
R&=&-\frac{D-1}{\kappa}\left(\frac{\kappa^{'}}{\kappa}\right)^{'}
-(D-1)\frac{(D-2)}{\kappa}\left(\frac{\kappa^{'}}{\kappa}\right)^2,\nn
R_{0\mu\nu\rho}R_0^{~~\mu\nu\rho}&=&
(D-1)\frac{3}{\kappa}\left[\left(\frac{\kappa^{'}}{2\kappa}\right)^{'}\right]^2,\nn
R_{j\mu\nu\rho}R_k^{~~\mu\nu\rho}&=&
-\delta_{jk}\frac{3}{\kappa}\left[\left(\frac{\kappa^{'}}{2\kappa}\right)^{'}\right]^2
-2\delta_{jk}\frac{D-2}{\kappa}\left(\frac{\kappa^{'}}{2\kappa}\right)^4,\nn
R_{\mu\nu\rho\sigma}R^{\mu\nu\rho\sigma}&=&
6\frac{D-1}{\kappa^2}\left[\left(\frac{\kappa^{'}}{2\kappa}\right)^{'}\right]^2
+2\frac{(D-1)(D-2)}{\kappa^2}\left(\frac{\kappa^{'}}{2\kappa}\right)^4.
\eea
Since $\tilde g_{\rho\sigma}-g_{\rho\sigma}=B\kappa\eta_{\rho\sigma}$, 
the functional derivatives are taken using the relation
\be
\int(\tilde g_{\rho\sigma}-g_{\rho\sigma})\frac{\delta(\cdot\cdot\cdot)}{\delta g_{\rho\sigma}}
=BD\int\kappa\frac{\delta(\cdot\cdot\cdot)}{\delta\kappa}.
\ee

\end{document}